\documentclass[twocolumn,final,10pt,journal,a4paper]{IEEEtran}

\usepackage[usenames,dvipsnames]{xcolor}
\usepackage{amsmath,amsthm}
\usepackage{epsfig,amsfonts}
\usepackage{graphicx,amssymb,amsmath}
\usepackage{color}
\usepackage{bm}
\usepackage{bbm}
\usepackage[small]{caption}
\usepackage{balance}
\ifCLASSOPTIONcompsoc
\usepackage[caption=false,font=normalsize,labelfont=sf,textfont=sf]{subfig}
\else
\usepackage[caption=false,font=footnotesize]{subfig}
\fi

\usepackage[nolist]{acronym}
\usepackage{mathtools}
\mathtoolsset{showonlyrefs}
\usepackage{lipsum} 

\newcommand{\dataset}{\mathcal{X}}
\newcommand{\testset}{\mathcal{T}}

\newcommand{\Pfa}{\mathsf{P_{fa}}}

\newcommand{\Pd}{\mathsf{P_{d}}}

\newcommand{\estef}{\textcolor{black}}
\newcommand{\er}{\textcolor{black}}

\begin{document}

\begin{acronym}
\acro{ML}{machine learning} 
\acro{AWGN}{additive white Gaussian noise}
\acro{NN}{neural network}
\acro{RF}{random forest}
\acro{BPSK}{binary phase shift keying}
\acro{SNR}{signal to noise ratio}
\acro{SINR}{signal to interference and noise ratio}
\acro{mMTC}{massive machine type communications}
\acro{ROC} {receiver operating characteristic}
\acro{IG} {information gain}
\acro{IoT} {internet of things}
\acro{SIC}{successive interference cancellation}
\acro{RA}{random access}
\end{acronym}

\title{Grant-Free Access: Machine Learning for Detection of Short Packets}

\author{
\vspace{.3em}
    \IEEEauthorblockN{Estefan\'ia Recayte, Andrea Munari, Federico Clazzer\\[.1em]
    \IEEEauthorblockA{Institute of Communications and Navigation, German Aerospace Center (DLR), We{\ss}ling, Germany\\
     Email:  \{Estefania.Recayte, Andrea.Munari, Federico.Clazzer\}@dlr.de}\\
 }
 \thanks{This work has been accepted for publication at IEEE ASMS\&SPSC 2020.}
\thanks{\copyright 2020 IEEE Personal use of this material is permitted. Permission
from IEEE must be obtained for all other uses, in any current or future media, including
reprinting /republishing this material for advertising or promotional purposes, creating new
collective works, for resale or redistribution to servers or lists, or reuse of any copyrighted
component of this work in other works.}
}

\maketitle
\vspace{.3em}

\thispagestyle{empty} \pagestyle{empty}

\begin{abstract}
In this paper, we explore the use of machine learning methods as an efficient alternative to correlation in performing packet detection. Targeting    satellite-based  massive machine type communications and internet of things scenarios, our focus is on \estef{a common channel} shared among a large number of terminals via a fully asynchronous ALOHA protocol to attempt delivery of short data packets. In this setup, we test the performance of two algorithms, neural networks and random forest, which are shown to provide substantial improvements over {traditional} techniques.  Excellent performance is demonstrated in terms of detection and false alarm probability also in the presence of collisions among user transmissions. The ability of machine learning to extract further information from incoming signals is also studied, discussing the possibility to classify detected preambles based on the level of interference they undergo.
\end{abstract}


\begin{IEEEkeywords}
machine learning, preamble detection, random access, grant free protocols, machine-type communications.
\end{IEEEkeywords}

\vspace{1.2em}

 \section{Introduction}\label{sec:Intro}

\IEEEPARstart{S}{mall} \estef{ data networks are steadily gaining momentum as a new approach for communications via Low Earth Orbit (LEO) satellite constellations. Characterised by the exchange of short information packets, this paradigm finds natural application in \ac{mMTC}, and has been embraced by international terrestrial standards (e.g., 3GPP, NB-IoT and LTE-M) as well as by commercial solutions in the satellite domain  (e.g., \cite{Orbcomm,Kepler}).} In this context, data are generated by a vast population of terminals that transmit in a sporadic and unpredictable fashion. 
Accordingly, grant-free schemes based on variations of the basic ALOHA \ac{RA} protocol \cite{aloha} are commonly employed, where users send information in a fully asynchronous and uncoordinated way over the shared bandwidth. 
\estef{This approach is especially suitable for satellite-based \ac{IoT} use-cases, allowing simple transmitter implementations that match well the capabilities of the battery-powered, low-complexity devices often encountered in these scenarios.}
The computational burden is instead shifted at the receiver, which has to attempt data retrieval from an incoming stream where the arrival time of information units is not known in advance and where packets may interfere with each other. 

From this standpoint, \emph{packet detection}, i.e., the ability to understand whether a transmission is present and to identify its start, is a  key enabler to achieve good performance.  Indeed, operating the system at an exceedingly high false alarm probability may trigger decoding procedures unnecessarily, increasing the receiver computation complexity. Conversely, missing or erroneously estimating the beginning of a data unit could severely impact the decoding process, especially when the receiver relies on \ac{SIC} to resolve collisions \cite{SICpaolini}.
In most practical implementations, detection is accomplished by correlating the incoming stream with a know preamble sequence, and by applying a threshold criterion to flag the presence of a packet. n Despite being sub-optimal \cite{massey:corr}\cite{ChianiSyn}, this solution is widely employed in view of its simplicity. Good performance is achieved in lightly loaded systems, yet a severe degradation is experienced in the presence of interference, especially when short preambles are employed. The definition of more effective approaches represents thus a fundamental challenge to support \ac{mMTC} in \estef{LEO constellations}. 

In this perspective, the lack of a known optimal solution for detection in grant-free access channels has recently drawn attention to the use of \ac{ML} as a promising alternative to correlation. 
Some initial and interesting results in this sense are provided in \cite{zorzi:ml}, where the authors focus on the random access channel (PRACH) of LTE systems. Considering the availability of multiple orthogonal sequences distributed among users and transmitted over an OFDM-based physical layer to ask for resource grants, authors apply \acp{NN} to estimate which preambles were transmitted as well as their multiplicity. Departing from this setting, and having in mind \ac{mMTC} applications in both the satellite and terrestrial domain, we tackle in this paper a fully uncoordinated scenario. Terminals do not undergo any negotiation procedure, but rather access the medium via an asynchronous ALOHA policy to  directly send a short data packet that includes few and common preamble symbols. In this setup, we explore the potential of two \ac{ML} algorithms, \acp{NN} and \ac{RF}, studying their performance in terms of detection and false alarm probabilities. Remarkable improvements over correlation-based methods are shown in a wide range of signal-to-noise ratio configurations, especially when transmissions are corrupted by mutual interference. Moreover, we suggest to leverage the  multi-classification capabilities of \ac{ML} to gather additional information on incoming signals. In particular, we consider the possibility to differentiate whether a preamble was received interference-free or endured a collision, involving one or more packets. Such an approach can be beneficial at the receiver, e.g. in order to determine the decoding order when \ac{SIC} is applied.
 
The remainder of the paper is organised as follows. We provide a brief and general overview of \ac{NN} and \ac{RF} in Sec.~\ref{sec:neuralnet}, followed in Sec.~\ref{sec:sysmodel} by a description of the system model considered in our study. Sec.~\ref{sec:algorithms} details the algorithms used for detection, whose performance is discussed in Sec.~\ref{sec:results}.
 
\section{Machine Learning Background} \label{sec:neuralnet}

The vast and complex nature of \ac{ML} renders a proper introduction to the topic well beyond the scope of this article. We provide instead a short high level overview of the two algorithms that will be later employed for packet detection, highlighting their key features in order to facilitate an understanding of the main results of our work. For a more comprehensive overview of the methods, the interested reader is referred to, e.g., \cite{montavon:nn, tin:RF, Simeone:ML}.

Throughout our discussion, we focus on supervised learning models for classification problems, which operate in two phases. During the first, taking place offline, the machine \emph{learns} by being fed with a set of inputs of known classes, i.e., each input is presented together with the corresponding classification label. Once the \emph{training} phase is complete, the algorithm can be run online, returning a classification prediction for any new unlabeled observation it is provided with. More formally, let ${\dataset = \{ \bm{x}_1, \bm{x}_2, ..., \bm{x}_d \}} $ indicate a training data-set composed by $d$ input vectors of $m$ elements each, also referred to as the number of features of the sample vector, and denote as \{$k_1, k_2, \dots, k_n\}$ the corresponding set of classification labels.
The performance of the trained \ac{ML} method can be measured by feeding a test data-set  ${\testset = \{ \bm{t}_1, \bm{t}_2, ..., \bm{t}_s \}}$, and checking the output predicted  labels \{$y_i$\}, $i=1,\dots,s$. Following this notation, the overall accuracy $\mathsf{A}$ of the algorithm is given by 
\begin{equation}\label{eq:acc}
  \mathsf{A} := \frac{1}{s} \sum_{i=1}^s \mathbbm{1} \{y_i= k_i\}  
\end{equation}
where  $\mathbbm{1}(\cdot)$ is the indicator function and note that the corresponding classifications labels $k_i$ of the test data-set are known. The metric offers thus an estimate of the probability to correctly label an incoming vector, computed based on the available training and testing sets. The definition can straightforwardly be extended to evaluate the ability of a trained \ac{ML} method to classify inputs belonging to specific classes of interest, as will be discussed in more details in Sec.~\ref{sec:algorithms}. 

Among the variety of supervised learning techniques, we focus on two widespread and efficient solutions, namely   \acp{NN} and \acp{RF}, which are   briefly reviewed in the following.

\subsection{Neural Network}
An artificial \ac{NN} \cite{montavon:nn} consists of a multipartite directed graph. Nodes, also called \emph{neurons}, are arranged in multiple distinct layers, so that a neuron belonging to a layer is connected to all and only neurons of the next layer through directed weighted edges. Each node in the first layer is associated with one element of an input vector, i.e., the number of neurons equals the cardinality of features for the considered input, whereas intermediate or \emph{hidden} layers may be composed by a variable number of nodes. Moreover, every neuron applies an activation function, combining the (weighted) inputs of the nodes of the previous layer and computing a scalar output, which is propagated forward through the outgoing edges. Finally, the last (\emph{output}) layer is populated by as many nodes as the possible number of classes, and computes the likelihood for the fed data to be classified with a specific label. 
  
The structure of the neural network is associated with a cost function, equal to zero for a perfect classification, and increasing as input vectors are wrongly labelled. In this sense, the aim of the training is to identify the appropriate weights that minimize the overall cost. Different algorithms to tackle the problem exist, among which a relevant role is played by the so called \emph{back-forward propagation}.
The method starts with a random initialization of the edge weights. Then, during a  forward propagation phase, the training data set is given as an input and the data is processed by the network to produce an output prediction. In the back propagation, the algorithm takes the predicted output given by the network,  compares it  with the correct classification and calculates an error function at the output layer. Then, the iterative gradient descent algorithm \cite{gradient:dis} is used to iteratively minimize the cost function by updates of the weight values. Once the last layer is adjusted, the algorithm back-calculates the error associated with the neurons edges from the preceding layer until the input layer is reached.  When the model is trained, the final structure of the neural network is used for predicting new unlabeled data. A new  observation is given as input and the NN gives the corresponding predicted classification as output.

\subsection{Random Forest}
Following a different approach to classification problems, \acp{RF} rely on the notion of binary decision trees. Within a decision tree, every node represents the evaluation of a feature for a provided input, its branches indicate the outcomes of such evaluation, and tree leaves are associated with the categorization labels. Each decision tree populating the forest is grown independently during the training phase following a top-down approach. Specifically, a random feature is chosen, and all vectors in the input data-set are split in two disjoint subsets by applying a threshold on the value they exhibit for the feature under analysis. The quality of the split is then evaluated by applying a criterion of interest, and the procedure is repeated considering other randomly chosen features. Once this is done, the feature leading to the split with the best quality (together with the employed threshold) is associated with the root node of the tree. The whole process is then repeated for the children nodes, building the tree in an iterative fashion.

In creating a forest, criteria to evaluate the obtained splits, as well as to stop the growth of trees, shall be provided. Common examples of the former are the Gini index and \ac{IG} \cite{forest:IG}. As to the latter criteria,  can be  for instance driven by a trade-off between accuracy and data overfitting or by defining a  threshold on  the splitting rule. When the stopping condition is fulfilled, each node without children is declared to be a leaf, and it is assigned the label of the category for which most input vectors reach it.

Finally, when a new observation has to be classified, the input sample is fed to each tree of forest. Here, the vector of features travels from the root downwards according the evaluated conditions until it reaches a leaf, which determines the prediction cast by the tree. As trees are grown independently, different classifications for the same observation sample may be obtained, and the \ac{RF} algorithm bases its final prediction on a majority voting criterion. 
 
 \section{System Model}\label{sec:sysmodel}

We focus on a scenario where a number of users access a shared wireless channel to send information to a single receiver in the form of short data packets. A simple ALOHA policy \cite{aloha} regulates access to the medium, with nodes transmitting data units in a fully asynchronous way. Such a setting is of particular interest for \ac{mMTC} {via satellite}, where large populations of terminals may sporadically generate and transmit small amount of information toward a sink in an unpredictable and uncoordinated fashion.
 
\begin{figure}[t]
 \includegraphics[width=0.25\textwidth]{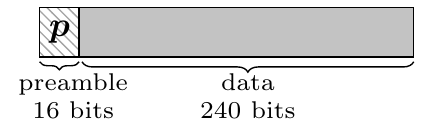}
 \centering \caption{Packet structure considered where data is preceded by a preamble sequence.}
 \label{fig:packet}
\end{figure}

As reported in Fig.~\ref{fig:packet}, packets are assumed to be overall $N= 256$ bits long, with the first $L=16$ bits reserved to a preamble (syncword) taking the form
\begin{equation}\label{preamble_seq}
\bm p = [1 1 1 0 \, 1 0 1 1 \,  1 0 0 1 \, 0 0 0 0 ].
\end{equation}    
The sequence $\bm p$ was originally proposed by the Consultative Committee for Space Data Systems, and specifically designed to perform well in detection of short packets thanks to good auto-correlation properties \cite{pre:CCSDS}. 

After \ac{BPSK} modulation, packets are sent over a channel with \ac{AWGN},  possibly interfering with each other. All users transmit with the same power level, and performance is tested for different \ac{SNR} levels at the receiver, namely $0$, $3$ and $8$ dB. For simplicity, users are assumed to be symbol-synchronous, i.e., no timing offsets are considered, and no frequency offset is present either.\footnote{ 
\er{In  practice,  symbol  synchronization  can be easily approached with the use of the oversampling technique. Furthermore, we remark that such an assumption does not alter the intrinsically asynchronous behaviour of the access protocol, regulated by a pure ALOHA policy.}}

\begin{figure*}[t!]
\centering
   \subfloat[]{\label{sm_awgn}
        \includegraphics[width=.4\textwidth]{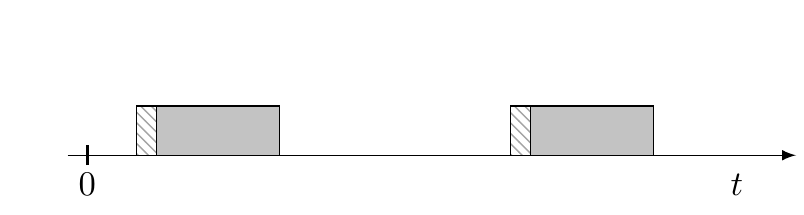}
    } \hspace{1em}
    \subfloat[]{\label{sm_interf}
        \includegraphics[width=.4\textwidth]{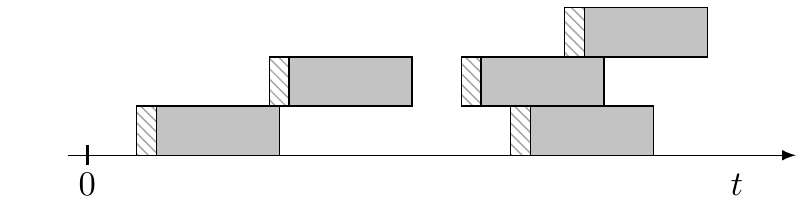}
    }      
\caption{Scenarios considered: (a) AWGN scenario with no collisions  and (b)  interference scenario where multiple transmissions occur.} \label{fig:scenarios}
\end{figure*}

To explore the capabilities of detection algorithms, two relevant scenarios are studied, summarised in Fig.~\ref{fig:scenarios}. First, we focus on purely \ac{AWGN} conditions, having receiver attempt detection when a single user is accessing the channel (Fig.~\ref{sm_awgn}). This setup offers a benchmark for the achievable performance, and can provide fundamental insights on the behaviour of different detection methods. Moreover, the obtained results can be representative for \ac{mMTC} grant-free systems when operated in lightly loaded conditions. We then complement our analysis considering the practically relevant case of a true interference channel, where multiple users contend for the medium
 (Fig.~\ref{sm_interf}). Since simultaneously transmissions may occur, the received signal can be corrupted by noise and interference, rendering detection far more challenging. 

In both scenarios, we evaluate the performance of detection algorithms by deriving false alarm probability $\Pfa$ and correct detection probability $\Pd$. Formally, we denote by $\mathsf{S}$ the event of having the \estef{complete preamble sequence}  within the observation window of length $L$, and by $\overline{\mathsf{S}}$ the complementary event, i.e. having a sequence which contains noise, data, or part of the syncword (or combinations thereof in case of interference). Furthermore, we indicate as $\mathsf{
D}$, the event of having the employed detection algorithm declare    the presence of the preamble sequence within the observed window. Following this notation, we define  
\begin{equation} \label{eq:pd_pfa}
\Pd := \Pr \left \{ \mathsf{
D}\, |  \, \mathsf{S} \right \}, \quad \Pfa := \Pr \left \{ \mathsf{
D}\, | \, \overline{\mathsf{S}} \right \}.
\end{equation}
From \eqref{eq:pd_pfa}, the performance of an ideal detector is achieved for $\Pd = 1$ and ${\Pfa = 0}$. \estef{Note that $\Pd$ and $\Pfa$ are not complementary events.}

\section{Detection Algorithms}
\label{sec:algorithms}

Detection performance is evaluated for three algorithms: correlation, neural networks and random forest.

\subsection{Correlation} By virtue of its simplicity, correlation represents the de facto solution for detection in most practical systems. In order to identify the start of a preamble sequence within an incoming symbol stream, the receiver operates via a sliding window of duration $L$ symbols, corresponding to the preamble length. In details, at symbol time $n$, the known preamble sequence $\bm p$ is correlated with the received stream, to obtain
\begin{equation}\label{eq:corr}
c(n)   = \sum_{i=0}^{L-1}  \bm{x}(i+n) \cdot \bm {p}(i)
\end{equation}
 where $c(n)$ indicates the correlation output,   $\bm {x}(j)$ is the incoming signal at time $j$ and $\bm p(i)$ is the $i-$th symbol of the preamble sequence defined in \eqref{preamble_seq}. The obtained value is then compared against a predefined threshold, and the start of a preamble is declared when $c(n)$ exceeds the value. The choice of the threshold triggers a key design trade-off, driving the balance between false alarm rate and probability of correct detection. It shall be noted that, despite its widespread use, correlation does not represent the optimal solution even under \ac{AWGN} conditions \cite{massey:corr}, and its performance in the presence of interference may be severely hindered.

\subsection{ML for packet detection}
From a \ac{ML} perspective,  packet detection can be treated as a classification problem, consisting in categorising input data onto a set of predefined classes. In the simplest instance, the input data can be represented by a portion of the incoming symbols stream, which needs to be classified as either being a preamble or not. Nonetheless, \ac{ML} offers further possibilities, allowing to process richer sets of information and to extract more in-depth classification outputs. For instance, the ability to determine whether an input vector contains at least a portion of a preamble, or to understand if a detected preamble is affected by interference or not, may trigger further improvements in the receiver design, as will be discussed in further details in Sec.~\ref{sec:results}. 

\noindent{\textbf{Performance metrics.}} As for any supervised learning model, the performance of both \ac{NN} and \ac{RF} applied to detection can be evaluated resorting to the accuracy metric $\mathsf{A}$  defined in \eqref{eq:acc}. In particular, let us denote without loss of generality as ${k_i=1}$ the label indicating  that the provided input sample vector $\bm{x_i}$ corresponds to the start of the packet (preamble sequence). Following this notation, the probability of false alarm $\Pfa$ defined in \eqref{eq:pd_pfa} can be estimated over a set of $s$ test samples as 
\begin{equation}
\Pfa \simeq \frac{\sum_{i=1}^s \mathbbm{1}\{y_i = 1, k_i \neq 1\}}{\sum_{i=1}^s \mathbbm{1}\{y_i = 1\}}
\end{equation}
where the numerator sums up the instances in which the machine erroneously classifies an input as a preamble, while the denominator accounts for all the fed samples that were predicted as a preamble.
Similarly, an estimate of the correct detection probability $\Pd$ for a \ac{ML} method can be derived as
\begin{equation}\label{eq:pd}
 \Pd \simeq \frac{\sum_{i=1}^s \mathbbm{1}\{y_i = 1, k_i = 1\}}{\sum_{i=1}^s \mathbbm{1}\{k_i = 1\}}.
\end{equation}

\noindent\textbf{ML implementation.} In order to evaluate detection and false alarm probability for both \ac{NN} and \ac{RF}, a training dataset $\dataset$ as well as a testing dataset $\testset$ with the corresponding set of classification labels were generated for each of the \ac{SNR} considered and for each scenario described in Sec.~\ref{sec:sysmodel}-\ref{fig:scenarios}.

Every dataset contains $10^4$ vectors, generated as follows. In the \ac{AWGN} case, a window of duration equal to twice the packet length is considered, and a single packet is randomly placed therein. Then, we extract five vectors of $L=16$ symbols each: one contains the preamble (corrupted by noise), while the others are randomly picked within the window duration. The process is then repeated until the whole dataset is filled. Similarly, in the interference scenario, three packets are placed uniformly at random within a window of $4\cdot N$ symbols (i.e., we consider a system operating at a channel load $\mathsf G = 0.75$ [pkt/pkt duration]). A total of $15$ $L$-symbol vectors are then taken, including the three preambles, prior to drawing a new realisation. 
 The sequences obtained in this way are then complemented by adding an additional feature taking the form
\begin{equation}\label{eq:seq}
 \bm{ x}_i = \Big[r_{l}, r_{l+1}, \dots , r_{l+L-1}, \sum_{k=0}^{L-1} r_{l+k}^2 \Big]. 
\end{equation}
Here, $l$ indicates the  position of the first symbol of the considered $L$-symbol sequence within the stream and the last feature corresponds to the aggregate power over the period, thus leading to $17$-component input vectors for the \ac{ML} algorithms.  Information about the power  can help in differentiating preambles from sequences with portions of noise-only samples, and was shown to offer a slight increase in performance by means of a dedicated study, whose outcome is not reported here due to space constraints. Additional parameters employed in the implementation of the \ac{ML} algorithms are summarized in Table~\ref{tab:ml}. \estef{The table parameters are  sub-optimized based on the greedy research algorithm.}

 \begin{table}[]
     \centering
     \begin{tabular}{l c}
        \hline
 \multicolumn{2}{c}{\textbf{Neural Network}} \\
  \hline			
  Number of hidden layers & 2  \\
  Neurons at input layer & 17 \\
  Neurons at first hidden layer & 325 \\
  Neurons at second hidden layer & 320 \\
  Training algorithm & back-forward propagation  \\
    & gradient descent \\[.3em] 
  \hline 
   \multicolumn{2}{c}{\textbf{Random Forest}} \\
  \hline
  Numbers of trees  & 100 \\
  Splitting algorithm & information gain (IG) \cite{gradient:dis}\\
  Stopping criteria   & \ac{IG} $< 0.01$ \\
  \hline 
  \hline\\
     \end{tabular}
     \caption{Parameters used for the considered \ac{ML} algorithms.}
     \label{tab:ml}
 \end{table}

\section{Results}\label{sec:results}
\er{ In this section, we present the results obtained for packet detection with the proposed \ac{ML} schemes in both the AWGN and interference scenarios. As a foreword, we observe that optimal metrics for the  synchronization issue, which is close to our detection problem, have been studied for the AWGN channel, for example in \cite{massey:corr} \cite{ChianiSyn}. However, such metrics  are not valid in interference-beset settings, for which performance bounds remain elusive.  In view of this, the sub-optimal yet practical and widely employed correlation technique has been chosen as reference for comparison.}

\subsection{\ac{AWGN} Scenario}
\begin{figure}[t]
\includegraphics[width=0.4\textwidth]{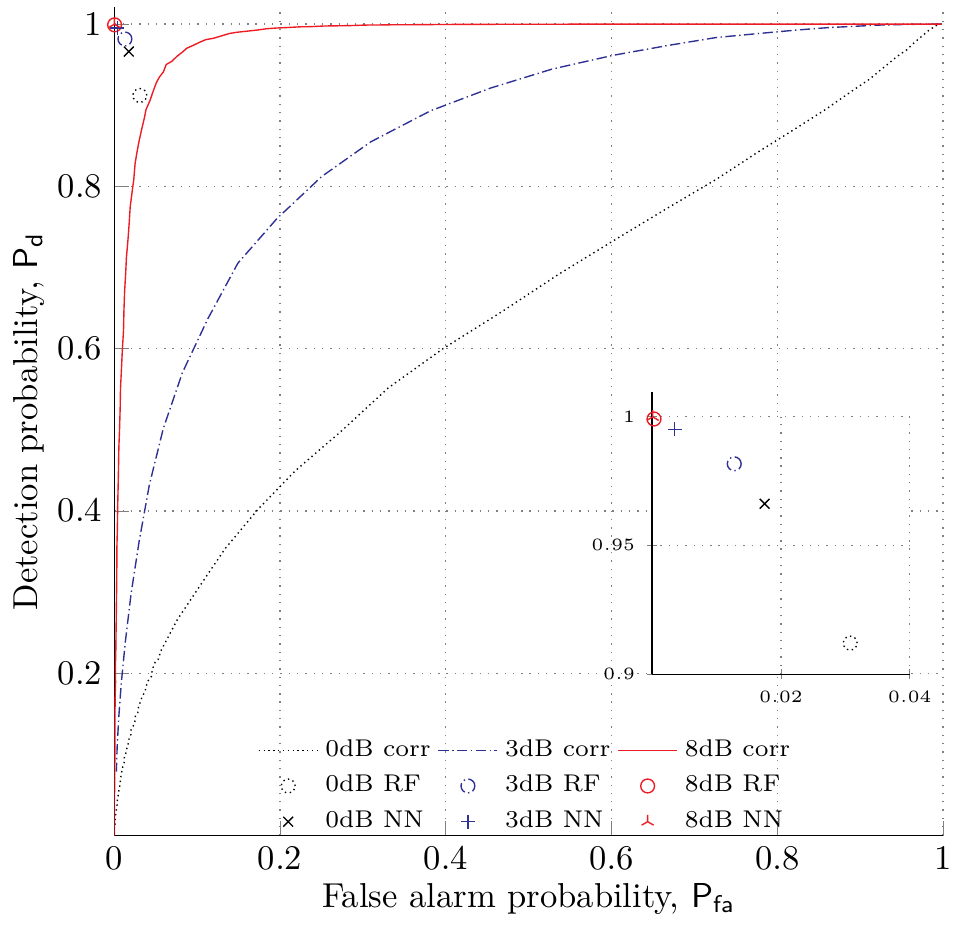}
 \centering \caption{Probability of detection in function of the false alarm considering a  \ac{AWGN} scenario for  ${\text{SNR}= 0, 3, 8}$ dB. Curves represents the results obtained with the correlation technique while circles and crosses represent the results obtained by neural network  and random forest, respectively.}
 \label{fig:awgn}
\end{figure}

We start our discussion focusing on an interference-free case. In this setting, Fig.~\ref{fig:awgn} reports the detection probability $\Pd$ against the false alarm probability $\Pfa$, i.e. the \ac{ROC} diagram. Let us first focus on the curves, which represent the performance achieved when using correlation. In this case, each point was obtained by changing the threshold value used to declare the presence of a preamble, and the plot clearly highlights the trade-off between high false alarm probabilities (obtained by setting low thresholds) and good detection. As expected, the behaviour improves for higher values of SNR. In particular, reasonable performance are obtained for a SNR of $8$ dB, where correlation can provide a \estef{$\Pd=0.96$ while granting $\Pfa \simeq  0.062$}. On the other hand, already when operating at $3$ dB, a detection probability larger than $90\%$ can only be achieved by tolerating a false alarm probability of more than $40\%$. This may unnecessarily trigger the decoding process very often, with detrimental effects on computational complexity and energy efficiency. 

Conversely, the plot reveals a remarkable potential in using supervised learning algorithms for small packet detection. The results for both \ac{ML} solutions, indicated by single points, consistently outperform correlation and, even at low levels of SNR, offer a high probability of detection while seldom triggering false alarm. Already at $0$ dB, a detection probability of $90\%$ is obtained with false alarms far lower than $5\%$. Moreover, Fig.~\ref{fig:awgn} clarifies how \ac{NN} (non-circle markers) offers better performance than \ac{RF} (circle markers), regardless of the SNR. This is confirmed when evaluating the results through the overall accuracy metric defined in \eqref{eq:acc} and considering a binary classification problem (i.e. preamble, non-preamble): at $0$ dB, we have ${\mathsf{A} = 0.9771}$ for  \ac{NN}, and ${ \mathsf{A} = 0.9502}$ when using \ac{RF}.
Such a trend emerges also in the other settings we considered, and highlights to the use of \ac{NN} as supervised learning scheme. From this standpoint, it shall be however noted that all the presented results were obtained by running the algorithms according to the parameters given in ${\text{Table \ref{tab:ml}}}$, which were not specifically optimised. While this further stresses the potential of \ac{ML}, leaving room for improvement, it also prompts the need for additional tuning and optimisations, in order to clarify which method (if any) shall be preferred. We regard these aspects as part of our future work.
 
Results presented so far were obtained employing \ac{ML} for binary classification. As discussed, however, the receiver can also benefit from additional information, gathered by defining further classes for the data-set. For instance, identifying that an input vector \--- although not corresponding to the start of a packet \--- contains a portion of the preamble followed by data, may allow to infer that a detection was missed, possibly triggering countermeasures. To explore the ability to classify data into multiple categories, both \ac{NN} and \ac{RF} were then tested considering six distinct labels: noise (n), noise and preamble (n-p), preamble (p), preamble and data (p-d), data (d), data and noise ({d}-n). \estef{Note that all labels refers to noisy observations. For instance, the difference between (n-p) and (p) is that the former corresponds to a sequence formed by    noise samples  followed by noisy preamble's symbols  while the latter corresponds to the observation of the preamble sequence which is noisy.  }  The results of this study are  illustrated in the confusion matrices given in Table \ref{tab:awgnconf}, operating at a SNR of $3$ dB. Within each matrix, the element in position $(i,j)$ corresponds to the estimated probability that the model classifies an input vector as belonging to the class associated with the $i$-th row given that the real label is the one associated with the ${j\text{-th}}$ column. 
Ideally, when the accuracy of a model is  $\mathsf{A}=1$, the confusion matrix is the identity matrix, providing  correct classification of every sample. Moreover, $\Pd$ and $\Pfa$ can easily be derived from a confusion matrix (in this scenario). For the labelling of  Table \ref{tab:awgnconf}, indicating as $a_{ij}$ the element in position $(i,j)$, we have $\Pd=a_{22}$ while $\Pfa = \sum_{j=1}^6 a_{2j} - a_{22}$.

In this case, we obtain $\mathsf{A}=0.6329  $ for the \ac{NN} and  for \ac{RF} $\mathsf{A}=0.6421$. Even though the accuracy is lower than in the binary classification due to the fact that the \ac{ML} has more possibilities of miss-classifications, the detection and false alarm probabilities remain excellent    estimated. For \ac{NN}  we obtain $\Pd = 0.9958 $ and  $\Pfa = 0.0198$ while for \ac{RF} we obtain $\Pd = 0.9616$  $\Pfa = 0.0438$. 
 
\begin{table}[]\centering
     \caption{Confusion matrices for \ac{AWGN}   at 3 dB. Labels indicate: (n) noise, (p) preamble, (n-p) noise and preamble, (p-d) preamble and data, (d) data, (d-n) data and noise.}
  \begin{tabular}{l c c c c c c }
  \hline
  \multicolumn{7}{c}{\textbf{Neural Network}} \\
  \hline			
         & n & p &  n-p  & p-d & d & d-n   \\
   \hline
 n &     \textbf{0.7956} &  0.0018&   0.1326   & 0.0054&    0.0060  &  0.2233 \\
 p &        0  & \textbf{0.9958}    &    0   & 0.0054   & 0.0072   & 0.0072 \\
 n+p &   0.1169  &       0  & \textbf{0.6929}&   0.0492 &   0.1692 &   0.1693 \\
 p+d &    0.0024 &        0   & 0.0120  & \textbf{0.6425}  &  0.3221  &  0.1813 \\
 d &   0.0006   & 0.0006  &  0.0738  &  0.1956 &   \textbf{0.3965}  &  0.1447 \\
 d+n &   0.0845 &   0.0018  &  0.0888  &  0.1020  &  0.0990  & \textbf{0.2743}\\
    \hline\\	
 \multicolumn{7}{c}{\textbf{Random Forest}} \\
 \hline		
         & n & p &  p-n  & p-d & d & d-n   \\
  \hline 
        n   & \textbf{0.8136} &	0.0066& 	0.1620 &	0.0126 &	0.0162 &	0.2215    \\
    p    &      0.0024  &   \textbf{0.9616} &    0.0018 &    0.0060 &    0.0174 &    0.0162
   \\
    n+p &   0.0815  &   0.0048   &  \textbf{0.6671 }&    0.0414   &  0.1321  &   0.0804  \\
    p+d &   0.0072  &   0.0054 &    0.0366  &  \textbf{0.6605}  &  0.3565   &  0.1639
   \\
    d  &   0.0042 &    0.0126   &  0.0804&     0.1938   &  \textbf{0.3818   }&  0.1501\\
    d+n &   0.0911 &    0.0090    & 0.0522   &  0.0858    & 0.0960   & \textbf{0.3679}\\
    \hline
  \hline \\
   \end{tabular}
    \label{tab:awgnconf}
\end{table}
 
\subsection{Interference scenario}

\begin{figure}[t!]
 \includegraphics[width=0.4\textwidth]{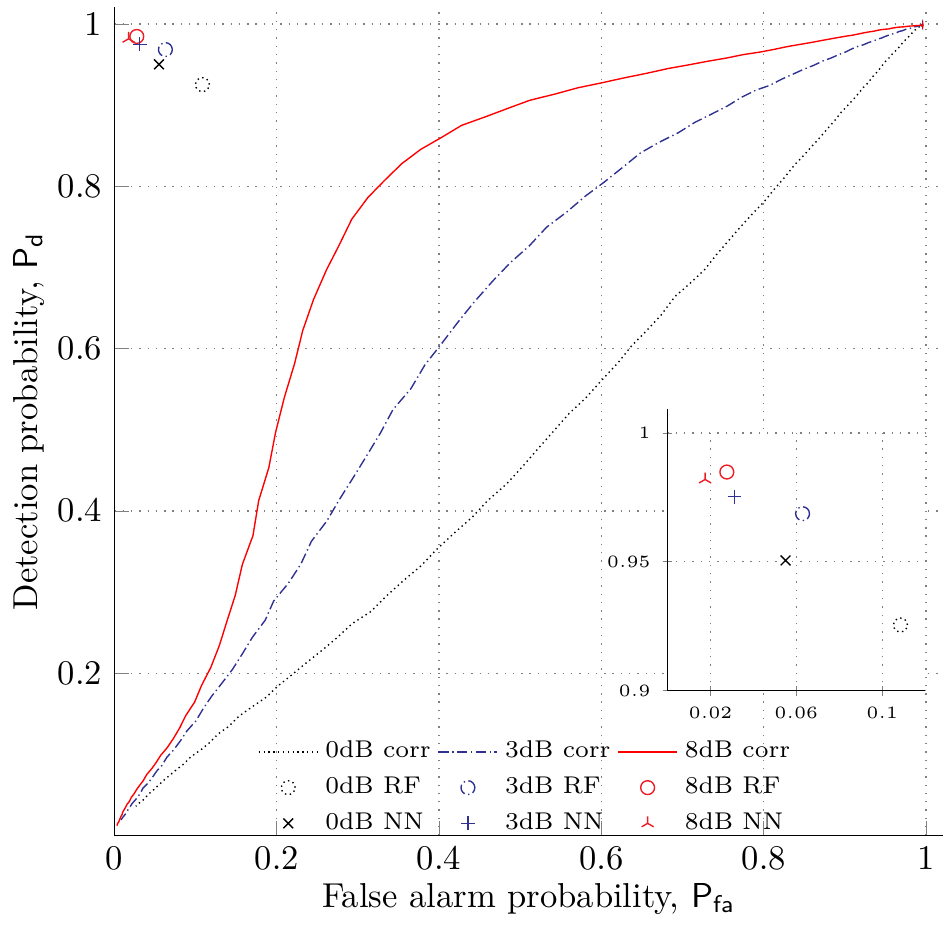}
 \centering \caption{Probability of detection in function of the false alarm considering a   interference scenario for  ${\text{SNR}= 0, 3, 8}$ dB. Curves represent the results obtained with the correlation technique while circles and crosses represent the results obtained by neural network  and random forest, respectively.}
 \label{fig:interf}
\end{figure}

Let us now focus on a scenario of practical relevance for transmission of small data packets for \ac{mMTC} applications, considering the case in which multiple users concurrently contend for the channel. Given the uncoordinated nature of the ALOHA medium sharing policy, simultaneous transmissions may lead to collisions, rendering the detection of packets corrupted by interference considerably more challenging. Fig.~\ref{fig:interf} illustrates the performance of the detection algorithms under study assuming a channel load $\mathsf{G} = 0.75$ [pkt/pkt duration], i.e on average 0.75 packets are received per packet duration. As expected, all methods present a performance degradation, due to the worse \ac{SINR} experienced at the receiver. The effect is especially strong for correlation, for which detection probabilities higher than $90\%$ come at the expense of a comparable level of false alarm probabilities, forcing the rest of the receiver chain to be activated almost at any incoming sample. Conversely,  both \ac{NN} and \ac{RF} are capable of offering an excellent detection behaviour even in the presence of multiple users, performing almost ideally for sufficiently large \ac{SNR} values. The achievable improvement can be for instance appreciated by observing how, if the receiver accepts a $\Pfa$ in the order of $10^{-2}$, the correlation technique reaches only a ${\Pd = 0.5734}$ at $8$ dB while with \ac{ML} a  ${\Pd = 0.98}$ can be ensured.

To conclude our study, we finally investigate the potential of multi-classification for \ac{ML} in a \ac{RA} setup. From this standpoint, the ability to understand whether a preamble is affected by a collision, or even to differentiate among different levels of interference, may enable important features to improve the overall system performance. For instance, in modern random access schemes where \ac{SIC} is employed \cite{Berioli16,Clazzer18}, the receiver may start the decoding process from interference-free data units, or determine the order of \ac{SIC} so as to tackle first less interfered packets, increasing the probability of success and possibly reducing the computational complexity of the process.

Having this in mind, four labels were specified for \ac{ML} to classify an input sample: no preamble (np), preamble (p), preamble with one interferer (p$_{+1}$), preamble with multiple interferers (p$_{+m}$). The confusion matrix obtained following this approach for a \ac{SNR} of $3$ dB are presented in Table~\ref{tab:interf}. 
 The \ac{NN} presents an accuracy $\mathsf{A} = 0.9335$ while a little degradation is given by the \ac{RF} model with  $\mathsf{A} = 0.8920$. The probabilities of detection, calculated according~\eqref{eq:pd}, are slightly lower than in the \ac{AWGN} scenario  $\Pd = 0.9714$ for \ac{NN} and $\Pd = 0.9311$ for \ac{RF} while the probability of false alarm for the interference scenario is similar to the one achieved in single-user scenario, i.e. $\Pfa = 0.0272$ for \ac{NN} and $\Pfa = 0.0324$ for \ac{RF}. More interestingly, the algorithms are capable to very accurately distinguish among interference-free preambles and preambles affected by a single interferer (elements along the main diagonal of the confusion matrix). Furthermore, taking \ac{NN} as reference, while preambles undergoing multiple collisions (p$_{+m}$) are correctly identified only $\sim 40\%$ of the times, in the vast majority of cases they are still perceived as being affected by interference, although from a single user (p$_{+1}$), proving the robustness of the algorithm.

\begin{table}[]
    \centering
       \caption{Confusion matrices for interference scenario at $3$ dB. Labels indicate: (np) not preamble, (p) preamble, (p$_{+1}$) preamble with one interferer, (p$_{+m}$) preamble with multi-interferers.}
    \begin{tabular}{l c c c c}
 \hline
  \multicolumn{5}{c}{\textbf{Neural Network}} \\
  \hline	
        & np & p & p$_{+1}$ & p$_{+m}$  \\
  \hline
    np  &  \textbf{0.9728}  &   0.0022 &   0.0520  &  0.1599 \\ 
    p   &   0.0026 &  \textbf{0.9696} &   0.0830  &  0.0093 \\ 
    p$_{+1}$ &   0.0204 &   0.0282 &  \textbf{0.8502} &  0.4504 \\ 
    p$_{+m}$ &   0.0042 &        0 &   0.0149  & \textbf{0.3804}\\
 \hline \\
    \multicolumn{5}{c}{\textbf{Random Forest}} \\
  \hline	
  & np & p & p$_{+1}$ & p$_{+m}$  \\
  \hline
    np    &\textbf{0.9676}  &  0.0213  &  0.1244  &  0.2107 \\
     p    & 0.0074   & \textbf{0.9396}&  0.1453  &  0.0220 \\
     p$_{+1}$  &  0.0202  &  0.0392  &  \textbf{0.7083}  &  0.5063 \\
      p$_{+m}$ &  0.0048  &       0  &  0.0220  &\textbf{0.2610}\\
  \hline
  \hline
  \\
    \end{tabular}
    \label{tab:interf}
\end{table}

\section{Conclusions}\label{sec:Conclusions}

{In this paper we showed the effectiveness of supervised learning techniques for short packet detection. We started by comparing the de facto correlation solution with artificial neural networks and random forest classifiers for a point-to-point link affected solely by \ac{AWGN}. The benchmark scenario already anticipates promising performance.  For instance, at $3$ dB and for a target false alarm probability $\Pfa$ of $0.003$ correlation provides a detection probability  {$\Pd$} \estef{$\simeq  0.11$} while \ac{RF} shows an improvement up to $0.995$. Having in focus an \ac{mMTC} scenario, we then moved to analyse an uncoordinated and asynchronous medium access setup with packet collisions. As envisaged the supervised learning detectors largely outperform traditional correlation. The \ac{NN} solution achieves up to $0.975$ detection probability for a target false alarm of $0.03$, an at least 10-fold improvement over correlation. Finally, we explored the multi-category classification capabilities, showing that the proposed machine learning detectors are also able to identify possible collisions, and eventually distinguish between one or more interferers, with reasonable accuracy. Such feature can be profitably exploited by a \ac{SIC} receiver so to process first lightly interfered packets having higher chances to be correctly decoded. \er{Relevant future work involves a computational complexity analysis of the ML schemes, together with a comparison with reference benchmarks that go beyond the correlation-based approach.}}

\balance
\bibliographystyle{IEEEtran}
\bibliography{references}

\end{document}